%% file: final.tex
\documentclass{ifacconf}

\usepackage{natbib}        
\usepackage{graphicx}
\usepackage{amsmath}
\usepackage{amssymb}
\usepackage{tikz,pgfplots}
\pgfplotsset{compat=newest}
\usetikzlibrary{patterns}
\usepackage{algorithm}
\usepackage{algorithmic}
\usepackage{subcaption}
\usepackage{caption}
\usetikzlibrary{patterns}
\usetikzlibrary{graphs}
\usetikzlibrary{graphs.standard}
\usetikzlibrary{matrix,arrows,fit,backgrounds,mindmap,plotmarks,decorations.pathreplacing}

\newtheorem{theorem}{Theorem}

\newtheorem{lemma}{Lemma}

\newtheorem{remark}{Remark}
\newtheorem{assumption}{Assumption}

\DeclareMathOperator{\1}{\textbf{1}}

\DeclareMathOperator*{\cov}{cov}
\DeclareMathOperator*{\diag}{diag}

\tikzstyle{sensor} = [draw, fill=blue!20, rectangle, rounded corners,
minimum height=2em, minimum width=7em]
\tikzstyle{est} = [draw, fill=orange!20, rectangle, rounded corners,
minimum height=2em, minimum width=7em]
\tikzstyle{pinstyle} = [pin edge={to-,thin,black}]
\begin{document}
\begin{frontmatter}

\title{Event-Based Control for Synchronization of Stochastic Linear Systems with Application to Distributed Estimation} 

\author[First]{Jiaqi Yan} 
\author[Second]{Yilin Mo} 
\author[First]{Hideaki Ishii}

\address[First]{Department of Computer Science, Tokyo Institute of Technology, Japan. Emails: {jyan@sc.dis.titech.ac.jp, ishii@c.titech.ac.jp}.}
\address[Second]{Department of Automation and BNRist, Tsinghua University, Beijing, China. E-mail: ylmo@tsinghua.edu.cn.}

\begin{abstract}                
This paper studies the synchronization of stochastic linear systems which are subject to a general class of noises, in the sense that the noises are bounded in covariance but might be correlated with the states of agents and among each other. We propose an event-based control protocol for achieving the synchronization among agents in the mean square sense and theoretically analyze the performance of it by using a stochastic Lyapunov function, where the stability of $c$-martingales is particularly developed to handle the challenges brought by the general model of noises and the event-triggering mechanism. The proposed event-based synchronization algorithm is then applied to solve the problem of distributed estimation in sensor network. Specifically, by losslessly decomposing the optimal Kalman filter, it is shown that the problem of distributed estimation can be resolved by using the algorithms designed for achieving the synchronization of stochastic linear systems. As such, an event-based distributed estimation algorithm is developed, where each sensor performs local filtering solely using its own measurement, together with the proposed event-based synchronization algorithm to fuse the local estimates of neighboring nodes. With the reduced communication frequency, the designed estimator is proved to be stable under the minimal requirements of network connectivity and collective system observability.
\end{abstract}

\end{frontmatter}

\section{Introduction}
Because of the widespread applications, the past decades have witnessed ever-growing interests in multi-agent systems (MASs). As one of its fundamental focuses, the synchronization problem for linear MASs has been studied in the literature. Given a group of agents modeled by general linear dynamics, the distributed controllers are proposed therein, towards the end of achieving the asymptotic consensus on the states of agents. For instance, in the continuous-time domain, \cite{ma2010necessary} find that minimal requirement for consensusability is that the agent dynamics has to be stabilizable and the communication graph must contain a spanning tree for unstable agent dynamics. On the other hand, \cite{you2011network} consider the MASs with discrete-time linear dynamics. Under a certain relation between the system matrix and the Laplacian matrix, the authors show that the synchronization among agents is able to be achieved with the static local feedback control. Since then, various control protocols are developed within this field of research, such as \cite{gu2011consensusability,xu2019distributed,yan2021resilient}.

On the other hand, the autonomous agents are often equipped with embedded microprocessors and onboard communication that are powered by energy-finite batteries and thus have limited energy resources. As such, event-triggering has recently been popular for their capabilities of improving the
resource utilization efficiency (\cite{dimarogonas2011distributed,yi2018dynamic,kadowaki2014event,nowzari2019event,yan2020resilient}). Since the early works \cite{garcia2014decentralized} and \cite{zhu2014event}, remarkable efforts have also been devoted to developing the event-based control protocols for solving the synchronization problem in deterministic linear MASs. In these years, stochastic linear systems with event-triggered schemes have received particular research attention. For example, \cite{ma2016event} focus on the stochastic MASs, where the dynamics of each agent is subject to mutually uncorrelated zero-mean Gaussian white noises. Using linear matrix inequalities, they provide an event-based controller which facilitates the synchronization among agents in the mean square sense while decreasing their communication frequency. Considering state-dependent noises, \cite{ding2015event} analyze control performance leveraging the theory of input-to-state stability in probability, and derive sufficient conditions under which the consensus in probability is reached by using an
event-triggered control protocol. 
However, despite these works, the results in this area have been scattered in literature.

In this work, we also focus on the event-based control for stochastic linear system synchronization. Different from the existing works, this paper considers a more general class of noises. It includes the independent Gaussian white noise model in \cite{ma2016event} and the state-dependent noise model in \cite{ding2015event}, since the noises are only assumed to be bounded in covariance while they might be correlated with the states of agents along time and among agents. Because of its generality, this problem reveals various applications in both theoretical and engineering fields (\cite{ding2015event,qi2020passivity}). 

Here, we follow the approach of \cite{2101.10689}, which finds that, by performing the local decomposition of the Kalman filter, the problem of distributed state estimation can be reformulated to that of the synchronization of stochastic linear MASs. We would therefore show how the event-based control protocol proposed in this work can be applied to develop stable distributed estimators as well. In particular, the aim is to estimate the state of an $n$-dimensional linear time invariant system by using the measurements from $m$ sensors, where the Kalman filter offers the optimal solution in a centralized manner (\cite{anderson2012optimal}). This paper presents a novel framework for the event-based distributed implementation of the centralized Kalman filter, where each sensor node performs local filtering with its own measurement based on a local decomposition of Kalman filter, and information fusion by communicating with neighbors through the proposed event-based synchronization algorithm. The designed estimator is proved to be stable (i.e., the state estimation error is mean-square bounded at each node side), which extends the results in \cite{2101.10689} for the full transmission case.

\section{Problem Formulation}\label{sec:problem}

Let us consider the synchronization of a group of $m$ agents. The dynamics of each agent $i$ is described by the following stochastic linear system:
\begin{equation}
	\eta_i(k+1) = S\eta_i(k)+Bu_i(k)+L_iz_i(k),
\end{equation} 
where $\eta_i(k)\in\mathbb{R}^n$ and $u_i(k)\in\mathbb{R}$ are respectively the state and control input, and $z_i(k)$ is the system noise with zero mean and bounded covariance, i.e., $\mathbb{E}(z_i(k))=0$ and $\cov(z_i(k))$ is bounded at any time. By assuming so, we consider the most general class of noises which might be correlated along time and among agents.

In this paper, we would like to design the control input for each node such that they are able to achieve synchronization in the presence of the noises $\{z_i(k)\}$. To this end, we make the following definitions:
\begin{align}\label{eqn:bareta}
	\bar{\eta}(k) \triangleq \frac{1}{m}\sum_{i=1}^m \eta_i(k),\; \bar L_z(k) \triangleq\frac{1}{m}\sum_{i=1}^m L_iz_i(k).
\end{align}
Namely, $\bar{\eta}(k)$ and $\bar{z}(k)$ respectively denote the averages of local states and noises at current time. Then the network of agents is said to reach synchronization in the mean square sense, if it holds at any $k\geq 0$ that:
\begin{enumerate}
	\item\textbf{Consistency:} The average of local states keeps consistent throughout the execution, i.e., 
	\begin{equation}\label{eqn:consistency}
		\bar{\eta}(k+1) =S\bar{\eta}(k)+\bar L_z(k).
	\end{equation}
	\item\textbf{Consensus:} The agents can achieve consensus with bounded error covariance, i.e., there exists a bounded $\Xi$ such that 
	\begin{equation}\label{eqn:consensus}
		\cov[(\eta_i(k)-\bar{\eta}(k))]\leq \Xi.
	\end{equation}
\end{enumerate}

We elucidate these statements as below. First, the consistency condition claims that the dynamics of average state $\bar{\eta}(k)$ is governed by a linear system, the input to which is the external noises only. Therefore, the interaction among agents will only affect the evolution of local states but not $\bar{\eta}(k)$. Second, due to the presence of uncertain noises, it is unable for the agents to achieve exact agreement, i.e., to perfectly track the average state $\bar{\eta}(k)$. As a compromise, in \eqref{eqn:consensus}, we aim to track $\bar{\eta}(k)$ with bounded error covariance. We should note that in the absence of noises, these conditions become identical to the standard ones as in \cite{you2011network,xu2019distributed,gu2011consensusability}. 

For a stable system matrix $S$, it is straightforward that simply holding the zero input, $u_i(k)=0$, can
realize the synchronization among agents. To avoid this trivial case, this paper focuses on the unstable system matrix:
\begin{assumption}
	At least one eigenvalue of $S$ lies on or outside the unit circle. Moreover, $(S,B)$ is controllable.
\end{assumption} 

The communication network is modeled by a connected undirected graph $\mathcal{G}=(\mathcal{V},\mathcal{E},\mathcal{A})$, where $\mathcal{V} =\{1,...,m\}$ is the set of agents, $\mathcal{E}\subset \mathcal{V}\times\mathcal{V}$ is the set of edges. Moreover, $\mathcal{A}=\left[a_{i j}\right]$ is the weighted adjacency matrix, where $a_{ij}\geq 0$ and $a_{ij}=a_{ji},\forall i,j \in \mathcal{V}$. Notice that $(i,j)\in \mathcal{E}$ if and only if $a_{ij}>0$. Let us arrange the eigenvalues of Laplacian matrix $\mathcal L_\mathcal{G}$ as 
\begin{equation}\label{eqn:mu}
0=\mu_1< \mu_2 \leq \cdots \leq \mu_m.
\end{equation}

\subsection{Event-triggered communication and control strategy}
This paper aims to design an event-driven strategy, under which synchronization among agents is reached with low communication frequency. In particular, for any agent $i$, it broadcasts to neighbors only when a certain event at its local side is triggered. Based on the received knowledge, agent $i$ updates its state according to
\begin{equation}\label{eqn:controller}
	u_i(k) =\Gamma \sum_{j=1}^m a_{ij}(\hat{\eta}_{j}(k)-\hat{\eta}_{i}(k)),
\end{equation}
where $\Gamma\in\mathbb{R}^{1\times n}$ is the matrix to be determined, and
\begin{equation}
	\hat{\eta}_{i}(k) \triangleq S^{(k-k_{s}^{i})}\eta_i(k_{s}^{i}), \; \; k\in [k_{s}^{i}, k_{s+1}^{i}),
\end{equation}
such that $\eta_i(k_{s}^{i})$ denotes the information most recently broadcast by agent $i$. 

To determine the triggering instants $k_{s}^{i}$ for agent $i$, let us define the triggering function $f_{i}(k)$ as
\begin{equation}\label{eqn:triggerfun}
	\begin{split}
		f_{i}(k)&=||\epsilon_{i}(k)||^2-(c_0+c_1\varrho^{k}),\\
	\end{split}
\end{equation}
where $c_0, c_1$ are positive constants, $\varrho\in(0,1)$, and
\begin{equation}\label{eqn:epsilon}
	\epsilon_{i}(k)\triangleq \hat{\eta}_i(k)-\eta_i(k).
\end{equation}
At the initial phase, communications at all agents are carried out. After that, each agent updates its local state based on \eqref{eqn:controller} until the triggering function \eqref{eqn:triggerfun} exceeds $0$. To be specific, once $f_{i}(k) \geq 0$, agent $i$ will be triggered. It then broadcasts its local state $\eta_{i}(k)$ to neighbors, yielding that $\epsilon_{i}(k)$ is reset to zero. Therefore, the sequence of triggering instants is determined recursively as
$	k_{s+1}^{i} \triangleq \min \left\{k>k_{s}^{i} \mid f_{i}(k) \geq 0\right\},\; k_{0}^{i}=0.$

\section{Performance Analysis}\label{sec:performance}
We shall next study the performance of the proposed algorithm. Particularly, in order to reach the synchronization among agents, it is desired to establish both consistency and consensus conditions. 

Notice that the presence of noises prevents us from directly applying approaches of Lyapunov stability for deterministic systems to the analysis. We therefore resort to a stochastic analogue of it. To this end, we will prove an important preliminary result for our purpose, i.e., $c$-martingale convergence lemma, which would be used later in the stability analysis:
\begin{lemma}[$c$-martingale convergence lemma]\label{lmm:cmartingale}
Let $\{\mathcal{F}(t)\}$ be a filtration in a
probability space $(\Omega,\mathcal{F},\mathcal{P})$. Moreover, let $\{V(k)\}$ be a sequence of non-negative random variables such that
	\begin{equation}\label{eqn:cmartingale}
		\mathbb{E}[\Delta V(k)] \leq -\rho V(k)+c(k),
	\end{equation}
where $\mathbb{E}[\Delta V(k)]\triangleq \mathbb{E}[V(k+1)|\mathcal{F}(k)]-V(k),
$ $\rho>0$, and $\mathbb{E}[c(k)]\leq c<\infty$. Then it holds for $k\geq 0$ that $\mathbb{E}\left[V(k)\right]$ is bounded.
\end{lemma}  
\begin{pf}
	It follows from \eqref{eqn:cmartingale} that
	\begin{equation}\label{eqn:cmartingale2}
		0\leq \mathbb{E}[V(k+1)|\mathcal{F}(k)]\leq (1-\rho)V(k)+c(k).
	\end{equation}
	The result is thus obvious by taking expectation on both sides of \eqref{eqn:cmartingale2}, which yields that
	\begin{equation}
		\begin{aligned}
			0&\leq \mathbb{E}\left[V(k+1)\right] \leq\left(1-\rho\right) \mathbb{E}\left[V(k)\right]+c\\
			&\leq \left(1-\rho\right)^{k+1} \mathbb{E}\left[V(0)\right]+c\sum_{t=0}^{k+1}\left(1-\rho\right)^t.
		\end{aligned}
	\end{equation}The proof is thus finished. \hfill$\square$
\end{pf} 

We are now ready to present the main result on achieving synchronization of stochastic linear systems:
\begin{theorem}\label{thm:synchronization}
	Suppose that the Mahler measure\footnote{The Mahler measure of a matrix is defined as the absolute product of unstable eigenvalues of it.} of $S$ meets the condition \begin{equation}\label{eqn:unstable}
		\prod_j |\lambda_j^u(S)| < \frac{1+\mu_2/\mu_m}{1-\mu_2/\mu_m},
	\end{equation}
	where $\lambda_j^u(S)$ denotes the $j$th unstable eigenvalue of $S$, and $\mu_2$
	and $\mu_m$ are defined in \eqref{eqn:mu}. Let $\Gamma$ be designed according to \begin{equation}\label{eqn:Gamma}
		\Gamma=\frac{2}{\mu_2+\mu_m}\frac{B^T\mathcal{P}S}{B^T\mathcal{P}B},
	\end{equation}
	where $\mathcal{P}>0$ is the solution to the following inequality
	\begin{equation}\label{eqn:riccati}
		\mathcal{P}-S^{T} \mathcal{P} S+\left(1-\zeta^{2}\right) \frac{S^{T} \mathcal{P} BB^{T} \mathcal{P} S}{B^{T} \mathcal{P} B}>0,
	\end{equation}
	with $\zeta$ satisfying $\prod_{j}\left|\lambda_{j}^{u}(S)\right|<\zeta^{-1} \leq\frac{1+\mu_{2} / \mu_{m}}{1-\mu_{2} / \mu_{m}}.$
	With the event-based controller \eqref{eqn:controller}, the network of agents reach synchronization in the mean square sense.. 
\end{theorem}

\section{Application to Event-based Distributed State Estimation}
There are many practices where the synchronization of stochastic linear systems is applicable. In this section, we will provide a non-trivial example showing how the results developed in the previous sections can be applied to solve the problem of distributed state estimation. 

To this end, let us first introduce the formulation of distributed estimation and consider the following discrete-time dynamical system:
\begin{equation}\label{eqn:plant}
	x(k+1) = Ax(k) + w(k),
\end{equation}
where $x(k)\in\mathbb{R}^n$ is the system state to be estimated, $w(k)\sim \mathcal{N} (0, Q )$ is the independent and identically distributed (i.i.d.) Gaussian disturbance with zero mean and covariance matrix $Q\geq 0$. The initial state $x(0)$ also follows the Gaussian distribution which has zero mean. 

A sensor network monitors the system above. The measurement from each sensor $i\in\{1,2,...,m\}$ is given by
\begin{equation}\label{eqn:sensoroutput}
	y_i(k) = C_ix(k) + v_i(k),
\end{equation}
where $y_i(k)\in\mathbb{R}$ is the measurement produced by sensor $i$, $C_i\in\mathbb{R}^{1\times n}$, and $v_i(k)\in\mathbb{R}$ is the measurement noise.

By collecting the measurements from all sensors, we have
\begin{equation}\label{eqn:sensormatrix}
	y(k) = Cx(k) + v(k),
\end{equation}
where
\begin{equation}
	\begin{split}
		y(k) \triangleq {\left[\begin{array}{c}
				y_{1}(k) \\
				\vdots \\
				y_{m}(k)
			\end{array}\right],} \; C \triangleq {\left[\begin{array}{c}
				C_{1} \\
				\vdots \\
				C_{m}
			\end{array}\right],} \;
		v(k) \triangleq {\left[\begin{array}{c}
				v_{1}(k) \\
				\vdots \\
				v_{m}(k)
			\end{array}\right]},
	\end{split}
\end{equation}
and $v(k)$ is zero-mean i.i.d. Gaussian noise with covariance $R\geq0$ and is independent of $w(k)$ and $x(0)$. 

In this section, we make the following assumption on system observability, which is the minimum requirement for solving the distributed estimation problem:
\begin{assumption}[Jointly observable]\label{assup:observable}
	The system is jointly observable, i.e., the pair $(A, C)$ is
	observable, while $(A, C_i)$ is not necessarily observable for each $i\in\{1,\cdots,m\}$.
\end{assumption}

\subsection{Fundamental limit: Kalman filter}
If there exists a single fusion center that has access to measurements from all sensors, then the centralized Kalman filter offers the optimal solution and provides a fundamental limit for all distributed estimation schemes. We thus introduce Kalman filter before going to the distributed solution.

Let $P(k)$ be the error covariance of Kalman estimate at time $k$. Under
Assumption~\ref{assup:observable}, the error covariance will converge to the steady state exponentially fast (\cite{anderson2012optimal}):
\begin{align}
	P=\lim _{k \rightarrow \infty} P(k). \label{eqn:KFcov}
\end{align}

Since a sensor network typically operates for a long period of time, we consider the steady-state Kalman filter, which has the fixed gain
\begin{align}\label{eqn:KFgain}
	K=P C^{T}\left(C P C^{T}+R\right)^{-1}.
\end{align}
By using $K$, the optimal Kalman estimate is calculated recursively as 
\begin{equation}\label{eqnn:optimalest}
	\begin{split}
		\hat{x}(k+1) =(A-KCA)\hat{x}(k)+Ky(k+1).
	\end{split}
\end{equation}

\subsection{An event-based distributed implementation of Kalman filter}\label{sec:decompose}
It is clear that the optimal estimate \eqref{eqnn:optimalest} requires the information from all sensors. However, in a distributed framework, each sensor is only capable of communicating with immediate neighbors, which renders the centralized solution impractical. Therefore, research attention has been devoted to approaching the performance of Kalman estimate in a distributed manner.


In this work, we propose a new framework for achieving the distributed estimation with an event-triggered communication strategy. Compared with existing algorithms, the novelty of this framework is that we decouple the procedure of local filters from that of the fusion process so that communication among sensors occurs only in the fusion process and will not affect the performance of local filters. Particularly, the update at each time can be divided into two phases. In the first phase, a lossless local decomposition of Kalman filter is proposed, based on which each sensor runs a local filter solely using its own measurement. After that, the second phase fuses the local knowledge by performing the event-based synchronization algorithm from Section~\ref{sec:problem}, to obtain the stable local estimators while decreasing the transmission frequency. 

In what follows, we will detail our solution for solving the distributed estimation problem by respectively introducing these two phases.

\subsubsection{Phase I: Implementation of local filters}
In the first phase, a local decomposition of the Kalman filter is proposed, where each sensor runs a local filter based on its own measurement without communicating with others. It will be proved that the Kalman estimate \eqref{eqnn:optimalest} is indeed a linear combination of the local filters. Therefore, by properly fusing the local filters in the second phase, it is possible to approach the performance of the Kalman filter.

Before going on, we make the following assumption:
\begin{assumption}\label{assup:eigenvalue}
	\quad
	\begin{enumerate}
		\item all the eigenvalues of $A-KCA$ are distinct;
		\item $A-KCA$ and $A$ do not have any common eigenvalues.
	\end{enumerate}
\end{assumption}

\begin{remark}
	In case that $A-KCA$ does not satisfy Assumption \ref{assup:eigenvalue}, we can always perturb $K$ to ensure the validity of these assumptions with little loss of the estimation performance. This is feasible as the system is observable and thus the eigenvalues of $A-KCA$ can be freely assigned.
\end{remark}

Under Assumption~\ref{assup:eigenvalue}, we can find a non singular matrix $V$, by which $A-KCA$ is diagonalized as
\begin{equation}\label{eqn:diagK}
	A-KCA =V\Lambda V^{-1},
\end{equation}
where $\Lambda\triangleq \diag(\lambda_1,\cdots,\lambda_n)$ such that $\lambda_1,\cdots,\lambda_n$ are the eigenvalues of $A-KCA$.

In this paper, we consider the following local filter, which is performed by each sensor $i$ solely based on its local measurements:
\begin{equation}\label{eqn:xi_z}
	\begin{split}
		z_i(k) &= y_i(k+1)-\beta^T\hat{\xi}_i(k),\\
		\hat{\xi}_i(k+1) &= S \hat{\xi}_i(k)+\1_n z_i(k), \;\forall i\in\{1,\cdots,m\},
	\end{split}
\end{equation}
where $\hat\xi_i(k)$ and $z_i(k)$ are respectively the state and output of the local filter, $\beta\in\mathbb{R}^n$ and $G_i$ solve the following equation: 
\begin{equation}\label{eqn:betadef}
	\begin{aligned}
		(G_i-\1_nC_i)A=\Lambda G_i,\; \beta^TG_i=C_iA,
	\end{aligned}
\end{equation}	
and 
\begin{equation}\label{eqn:S}
	S \triangleq \Lambda + \1_n \beta^T.
\end{equation} 

Before analyzing the performance of the local filter \eqref{eqn:xi_z}, some of its properties are summarized below:
\begin{lemma}\label{lmm:beta}
Consider the local filter \eqref{eqn:xi_z}. There always exist $\beta$ and $G_i$ that solve \eqref{eqn:betadef}. Moreover, it holds for $k\geq 0$ that the covariance of $z_i(k)$ is bounded.
\end{lemma}

We shall next study the performance of local filter \eqref{eqn:xi_z}, namely, how it is related to that of the Kalman filter. For simplicity, decompose the Kalman gain as $K = [K_1,\cdots,K_m]$, where $K_j$ denote the $j$th column of $K$. In the next lemma, it is shown that the Kalman estimate \eqref{eqnn:optimalest} is indeed a weighted sum of the local filters and thus can be perfectly recovered by the collective information of them:
\begin{lemma}[\hspace{1pt}\cite{mo2016secure}]
	\label{lmm:decompose}
	Suppose Assumption \ref{assup:eigenvalue} holds. The Kalman estimate \eqref{eqnn:optimalest} can be recovered by $\hat \xi_i(k), i=1,...,m$, through a linear combination as below:
	\begin{equation}\label{eqn:localdecompose}
		\hat{x}(k) = \sum_{i=1}^{m}F_i \hat\xi_i(k),
	\end{equation}
	where\footnote{Notice $\diag (V^{-1}K_i)$ is a diagonal matrix, where the $j$th diagonal entry of $\diag (V^{-1}K_i)$ equals the $j$th entry of vector $V^{-1}K_i$.} $F_i = V \diag (V^{-1}K_i)$. 
\end{lemma}



\subsubsection{Phase II: Fusion of local filters via event-based synchronization algorithm}\label{sec:algorithm}
By virtue of \eqref{eqn:localdecompose}, it is clear that the optimal estimate needs to fuse $\hat{\xi}_j(k)$ of all sensors. However, in the distributed formulation, a single sensor can only access the local states within its neighborhood. Therefore, in the second phase, each sensor $i$ runs a synchronization algorithm, through which a stable local estimate is obtained by inferring $\hat{\xi}_j(k)$ for all $j\in\mathcal{V}$. For the particular purpose of decreasing the transmission frequency, we will adopt the event-based synchronization algorithm proposed in Section~\ref{sec:problem}.

To simplify notations, let $\eta_{i, j}(k)$ denote the inference from sensor $i$ on $\hat{\xi}_j(k)$.
By collecting its inference on all sensors, each sensor $i$ contains a local state
\begin{equation}
	\eta_i(k) \triangleq
	{\left[\begin{array}{c}
			\eta_{i, 1}(k) \\
			\vdots \\
			\eta_{i, m}(k)
		\end{array}\right]\in\mathbb{R}^{mn}.} 
\end{equation}
Particularly, the local inference will be updated through the following synchronization algorithm:
\begin{equation}\label{eqn:update}
	\eta_i(k+1) =\tilde{S}\eta_i(k) +\tilde{L}_iz_i(k)+\tilde{B}\tilde{\Gamma}\sum_{j=1}^m a_{ij}(\hat{\eta}_j(k)-\hat{\eta}_i(k)),
\end{equation}
where $\tilde{S} \triangleq I_{m} \otimes S, \;\tilde{L}_{i} \triangleq e_{i} \otimes \1_n$ with $e_{i}$ denoting the $i$th canonical basis vector in $\mathbb{R}^m$, $\tilde{B} \triangleq I_m \otimes B$ such that $(S,B)$ is controllable, $\tilde{\Gamma}=I_m\otimes\Gamma$, and $\hat{\eta}_i(k)$ is the latest information broadcast by sensor $i$ and is calculated by
\begin{equation}\label{eqn:hateta}
	\hat{\eta}_i(k) = \tilde{S}^{(k-k_{s}^{i})}\eta_i(k_{s}^{i}).
\end{equation}

\begin{figure}
	\centering
	\resizebox{0.4\textwidth}{!}{\input{tikz/eventblkdiagconf.tikz}}
	\caption{The information flow of Algorithm \ref{alg:dist_est}.}
	\label{fig:blkdiag}
\end{figure}
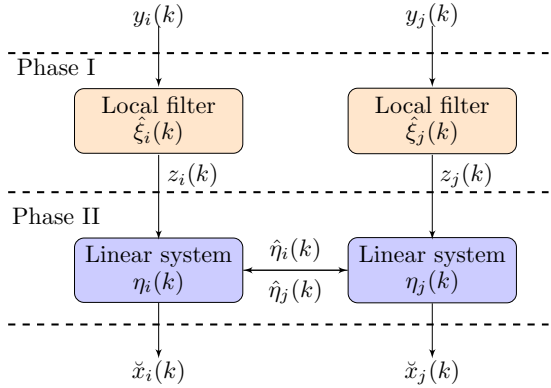

By collecting Phases I and II together, we summarize the update of any sensor $i$ in Algorithm~\ref{alg:dist_est}, where the initial values are set as $\hat{x}_i(0)=0$ and $\eta_i(0)=0$. The information flow of Algorithm~\ref{alg:dist_est} is provided in Fig.~\ref{fig:blkdiag}, which requires no fusion center and thus is achieved in a distributed manner.

\begin{algorithm}
	1:\: Sorely using its own measurement, sensor $i$ computes $z_i(k)$ and updates the state of the local filter by \eqref{eqn:xi_z}.\\
	2:\: By fusing the most recently received information within its neighborhood, sensor $i$ updates $\eta_i(k+1)$ according to the synchronization algorithm \eqref{eqn:update}--\eqref{eqn:hateta} and \eqref{eqn:triggerfun}--\eqref{eqn:epsilon}.\\
	3:\: Sensor $i$ updates the local estimate as
	\begin{equation}\label{eqn:localfuse}
		\breve{x}_i(k+1)=mF\eta_i(k+1),
	\end{equation}
where $F \triangleq\left[F_{1}, F_{2}, \cdots, F_{m}\right]$. \\
	4:\: Sensor $i$ broadcasts the new state $\eta_i(k+1)$ to neighbors if and only if the triggering function \eqref{eqn:triggerfun} exceeds $0$.
	\caption{An event-based distributed estimation algorithm for sensor $i$ at time $k>0$}
	\label{alg:dist_est}
\end{algorithm}

\subsection{Stability analysis of local estimators}
As shown in Lemma~\ref{lmm:beta}, the covariance of $z_i(k)$ in \eqref{eqn:update} is bounded at any time. As proved in Section~\ref{sec:performance}, the synchronization algorithm \eqref{eqn:update} guarantees that the local states $\eta_i(k)$'s achieve both the consistency and consensus conditions. We shall, in the sequel of this part, show how these conditions will help to achieve a stable local estimate at each sensor side.

\begin{theorem}\label{thm:general}
	Suppose that the Mahler measure of $S$ meets condition \eqref{eqn:unstable}, and $\Gamma$ is designed based on \eqref{eqn:Gamma} and \eqref{eqn:riccati}. Algorithm~\ref{alg:dist_est} yields a stable estimator at each sensor side. Specifically, it holds for any $k\geq 0$ that:
	\begin{enumerate}
		\item The average of local estimates from all sensor equals the Kalman estimate. That is,
		\begin{equation}
			\frac{1}{m}\sum_{i=1}^m \breve{x}_i(k)=\hat{x}(k), \forall k\geq 0.
		\end{equation}
		\item The error covariance of local estimate from each sensor $i$, i.e., $\cov(\breve{x}_i(k)-x(k))$, is bounded. 
	\end{enumerate} 
\end{theorem}

\begin{pf}
The proof respectively hold by the consistency condition \eqref{eqn:consistency} and consensus condition \eqref{eqn:consensus}. 

(1) Due to consistency condition \eqref{eqn:consistency}, we obtain
\begin{equation}\label{eqn:etasum}
\sum_{i=1}^m \eta_i(k+1) =\tilde{S}\sum_{i=1}^m\eta_i(k) +\sum_{i=1}^m\tilde{L}_iz_i(k).
\end{equation}
Comparing it with \eqref{eqn:xi_z}, we conclude the following relation for any time $k$ and any $j\in\mathcal{V}$:
\begin{equation}\label{eqn:xvseta}
\hat{\xi}_j(k) = \sum_{i=1}^m \eta_{i, j}(k).
\end{equation}
Thereby, the following equation is satisfied at any $k\geq 0$:
\begin{equation}
\begin{split}
\frac{1}{m}\sum_{i=1}^m \breve{x}_i(k)&=\sum_{i=1}^m F\eta_i(k)=\sum_{i=1}^m\sum_{j=1}^m F_j\eta_{i, j}(k)\\&=\sum_{j=1}^m F_j \Big[\sum_{i=1}^m \eta_{i, j}(k)\Big]=\sum_{j=1}^m F_j\hat{\xi}_j(k)=\hat{x}(k).
\end{split}
\end{equation}
(2) Let us consider the local estimator of any sensor $i$. To begin with, by the virtue of \eqref{eqn:consensus}, we conclude that $\cov(\delta_i(k))$ is bounded at any time $k$, where
\begin{equation*}
\begin{aligned}
\delta_i(k)=\eta_i(k)-\bar{\eta}(k).
\end{aligned}
\end{equation*}

In order to prove the boundedness of error covariance at each sensor side, let us denote
\begin{equation}
\bar{e}_i(k) \triangleq \breve{x}_i(k)-\hat{x}(k),
\end{equation}
which is the bias from local estimate $\breve{x}_i(k)$ to the optimal Kalman filter. Combining \eqref{eqn:localdecompose} and \eqref{eqn:xvseta} yields
\begin{equation}
\hat{x}(k) = F\sum_{i=1}^m \eta_i(k)= mF\bar{\eta}(k).
\end{equation}
One thus has
\begin{equation}\label{eqn:bar_e}
\bar{e}_i(k) = mF(\eta_i(k)-\bar{\eta}(k))=mF\delta_i(k).
\end{equation}
Therefore, the estimation error of sensor $i$ is calculated as
\begin{equation*}
\begin{split}
\breve{e}_i(k) &= \breve{x}_i(k)-x(k)=(\breve{x}_i(k)-\hat{x}(k))+(\hat{x}(k)-x(k)) \\&= \bar{e}_i(k)+\hat{e}(k),
\end{split}
\end{equation*}
where $\hat{e}(k)$ is the estimation error of Kalman filter. Since Kalman filter is optimal, $\bar{e}_i(k)$ is orthogonal to $\hat{e}(k)$. Therefore, it follows that
\begin{equation}\label{eqn:error}
\begin{aligned}
\cov(\breve{e}_i(k)) &= \cov(\bar{e}_i(k))+\cov(\hat{e}(k))\\&=m^2F\cov(\delta_i(k))F^T+P\leq m^2F\Xi F^T+P,
\end{aligned}
\end{equation}
where the inequality holds by the consensus condition \eqref{eqn:consensus}, and $P$ is steady-state error convariance of Kalman filter as defined in \eqref{eqn:KFcov}. We therefore complete the proof.\hfill$\square$
\end{pf}

Therefore, by applying \eqref{eqn:update}, the problem of distributed estimation is solved.

\section{Numerical Example}

In the numerical example, we consider the case where a network of four sensors monitors the system with following parameters:
\begin{equation}
	\begin{split}
		& A = 
		\begin{bmatrix}
			0.9 & 0\\
			0 & 1.1
		\end{bmatrix},\;
		C = 
		\begin{bmatrix}
			1 & 0 & 1 & 1\\
			0 & 1 & 1 & -1
		\end{bmatrix}^T,\;Q=0.5I_2, \;R=2I_4.
	\end{split}
\end{equation}

Suppose that the four sensors are connected as a ring, where each edge is assigned with weight $1$. 
The Laplacian matrix is thus:
\begin{equation}
\mathcal{L_G}=\begin{bmatrix}
2 & -1 & 0 & -1\\
-1 & 2 & -1 & 0\\
0 & -1 & 2 & -1\\
-1& 0 & -1 & 2
\end{bmatrix}.
\end{equation}
Therefore, one has
$\mu_2=2$, $\mu_4=4$. To guarantee the condition in Theorem~\ref{thm:general}, let us choose $\zeta=0.5$. Based on these parameters, we calculate $\Gamma=[0.80,-0.41].$

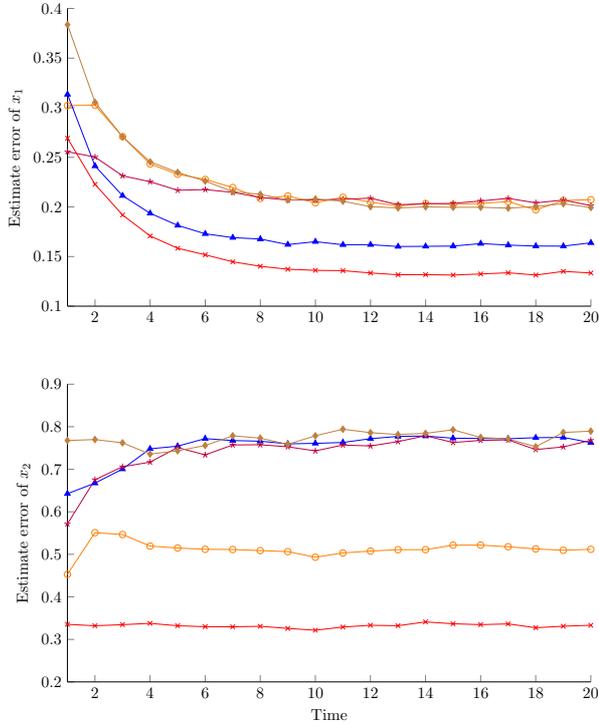
\begin{figure}[htbp]
    \centering
		\resizebox{0.5\textwidth}{!}{\input{tikz/eventx.tikz}}
	\caption{Average mean square estimation error of system states in $1000$-run Monte Carlo trials.}
	\label{fig:x}
\end{figure}

The initial states are set as $x(0)\sim \mathcal{N}(0,I)$ and $\breve x_i(0)=0$ for each sensor $i$. By performing Algorithm~\ref{alg:dist_est}, we observe from Fig.~\ref{fig:x} that the mean squared estimation error from each sensor is stable during the operation. Moreover, the average communication rate over the whole network is
$66.5\%$ obtained by the $1000$-run Monte Carlo trials. On the other hand, the performance loss is only $5.3\%$ compared to the full transmission case.


\section{Conclusion}
In this paper, the event-triggered synchronization problem has been addressed for the discrete-time stochastic MASs which are subject to a general class of noises. By using the stochastic Lyapunov stability theory, we propose an event-based control protocol, which is proved to facilitate the mean-squared synchronization among agents. In particular, through the local decomposition of the Kalman filter, we transform the problem of distributed estimation to the synchronization of stochastic linear systems, in which the proposed event-based protocol is applied to yield stable local estimators while decreasing the communication frequency.

\bibliography{reference}

\end{document}

%% file: tikz/eventblkdiagconf.tikz
\begin{tikzpicture}[auto, node distance=1.8cm,>=latex']
	
	\node at (0,3.75) {$y_i(k)$};
	\node at (4,3.75) {$y_j(k)$};
	
	\draw [->] (0,3.6) -- (0,2.7);
	\draw [->] (4,3.6) -- (4,2.7);
	
	\node [sensor,align=center] (est1) {Linear system\\$\eta_{i}(k)$};
	\node [sensor, right of=est1,node distance=4cm,align=center] (est2) {Linear system\\$\eta_{j}(k)$};
	\node [est,above of=est1,align=center,node distance=2.2cm] (sensor1) {Local filter\\$\hat{\xi}_i(k)$};
	\node [est,above of=est2,align=center,node distance=2.2cm] (sensor2) {Local filter\\$\hat{\xi}_j(k)$};

	\draw [->] (0,1.7) --  (0,0.45);
	\draw [->] (4,1.7) -- (4,0.45);
	
	\node at (0.5,1.4) {$z_i(k)$};
	\node at (4.5,1.4) {$z_j(k)$};
	
	\draw [->] (est1) -- node {$\hat\eta_i(k)$} (est2);
	\draw [->] (est2) -- node {$\hat\eta_j(k)$} (est1);
	
	
	\draw[dashed, line width=0.3mm] (-2.2, 3.2) -- (5.7, 3.2);
	\draw[dashed, line width=0.3mm] (-2.2, 1.15) -- (5.7, 1.15);
	\draw[dashed, line width=0.3mm] (-2.2, -0.8) -- (5.7, -0.8);
	
	\node at (-1.5,3.0) {Phase I};
	\node at (-1.5,0.8) {Phase II};
	
	\draw [->] (est1) --  (0,-1.3);
	\draw [->] (est2) -- (4,-1.3);

	\node at (0,-1.5) {$\breve{x}_i(k)$};
	\node at (4,-1.5) {$\breve{x}_j(k)$};
\end{tikzpicture}

%% file: tikz/eventx.tikz
%
%
\definecolor{mycolor1}{rgb}{0.00000,0.44700,0.74100}%
\definecolor{mycolor2}{rgb}{0.85000,0.32500,0.09800}%
\definecolor{mycolor3}{rgb}{0.92900,0.69400,0.12500}%
\definecolor{mycolor4}{rgb}{0.49400,0.18400,0.55600}%
\definecolor{mycolor5}{rgb}{0.46600,0.67400,0.18800}%
\begin{tikzpicture}

\begin{axis}[%
width=4.521in,
height=2.593in,
at={(0.758in,3.554in)},
scale only axis,
xmin=1,
xmax=20,
ymin=0.1,
ymax=0.4,
ylabel={Estimate error of $x_1$},
axis background/.style={fill=white},
axis x line*=bottom,
axis y line*=left
]
\addplot[color={orange}, mark={o}]
table[row sep=crcr]{%
1	0.302121180752325\\
2	0.302531925382281\\
3	0.270790892929535\\
4	0.243276799241696\\
5	0.232888346243705\\
6	0.227708282124915\\
7	0.219562708018138\\
8	0.208534774177071\\
9	0.211053634882657\\
10	0.204244486403274\\
11	0.209712530724859\\
12	0.204767961390524\\
13	0.20101659775227\\
14	0.203540896279591\\
15	0.202400557591565\\
16	0.203549485974262\\
17	0.205386769064392\\
18	0.19706440684058\\
19	0.20651303875293\\
20	0.207283075552277\\
};
\addplot[color={blue}, mark={triangle*}]
table[row sep=crcr]{%
1	0.313395924557476\\
2	0.241235028228442\\
3	0.211318017339914\\
4	0.193582756586918\\
5	0.181398105722853\\
6	0.172939605034201\\
7	0.169074571838835\\
8	0.167658191774771\\
9	0.162053809352254\\
10	0.165007345486911\\
11	0.161879212916129\\
12	0.161951350345411\\
13	0.160099139842983\\
14	0.160295758177786\\
15	0.160554931811933\\
16	0.163059177764199\\
17	0.161532948376557\\
18	0.16056133123955\\
19	0.160526809138623\\
20	0.163817507865157\\
};
\addplot[color={purple}, mark={star}]
table[row sep=crcr]{%
1	0.255628916441544\\
2	0.250275356313448\\
3	0.231336405929384\\
4	0.225384695644437\\
5	0.216804303447242\\
6	0.217448469635608\\
7	0.214714453980911\\
8	0.209494214895698\\
9	0.206950355341257\\
10	0.207044763187027\\
11	0.207671198596366\\
12	0.208831784760228\\
13	0.202192460601073\\
14	0.203233278598226\\
15	0.203549073895409\\
16	0.206022226570817\\
17	0.208622191746696\\
18	0.204092133426298\\
19	0.206947620037054\\
20	0.20132985249283\\
};
\addplot[color={brown}, mark={diamond*}]
table[row sep=crcr]{%
1	0.383593560200717\\
2	0.305409921380758\\
3	0.270690727454749\\
4	0.245413374750699\\
5	0.234761145971102\\
6	0.225984523706989\\
7	0.21538058064133\\
8	0.212691680438308\\
9	0.20697878705572\\
10	0.207942236185892\\
11	0.205738125212475\\
12	0.200243252998125\\
13	0.199030670595392\\
14	0.20004087672674\\
15	0.199644087660341\\
16	0.199642238130119\\
17	0.198791611562973\\
18	0.20033477136046\\
19	0.203273667574249\\
20	0.199362563701373\\
};
\addplot[color={red}, mark={x}]
table[row sep=crcr]{%
1	0.269151387318479\\
2	0.222787204893859\\
3	0.191867442683382\\
4	0.17071687357105\\
5	0.158362032512998\\
6	0.151740557392187\\
7	0.144694752495978\\
8	0.140171075377349\\
9	0.137224529367621\\
10	0.136089865134629\\
11	0.135768715821108\\
12	0.133440549024216\\
13	0.131664391376833\\
14	0.131750764538463\\
15	0.131352076748205\\
16	0.132451174818388\\
17	0.133625149467611\\
18	0.131341925245552\\
19	0.135041594288518\\
20	0.133428591892059\\
};
\end{axis}

\begin{axis}[%
	width=4.521in,
	height=2.593in,
	at={(0.758in,0.281in)},
	scale only axis,
	xmin=1,
	xmax=20,
	xlabel={Time},
	ymin=0.2,
	ymax=0.9,
	ylabel={Estimate error of $x_2$},
	axis background/.style={fill=white},
	axis x line*=bottom,
	axis y line*=left,
	legend style={at={(1.063,1.8)}, anchor=south west, legend cell align=left, align=left, draw=white!15!black}
	]
\addplot[color={orange}, mark={o}]
table[row sep=crcr]{%
1	0.453029400067562\\
2	0.550894916164929\\
3	0.546583406464047\\
4	0.519424517870761\\
5	0.514885188224169\\
6	0.51180799840207\\
7	0.511305994808284\\
8	0.508836527397316\\
9	0.506342619585707\\
10	0.493212400205813\\
11	0.503271154377489\\
12	0.507700269914462\\
13	0.51092684860295\\
14	0.510913991263884\\
15	0.521600196367656\\
16	0.521611587238072\\
17	0.517898687406905\\
18	0.512692699662478\\
19	0.509555786187111\\
20	0.511801907551975\\
};
\addlegendentry {{\Large s1}}
\addplot[color={blue}, mark={triangle*}]
table[row sep=crcr]{%
1	0.64256224120478\\
2	0.667407709595111\\
3	0.700505363251635\\
4	0.748112559935401\\
5	0.75429870078848\\
6	0.772128134221469\\
7	0.76752966610019\\
8	0.766036084665164\\
9	0.759784539467824\\
10	0.76093452019309\\
11	0.763007790740956\\
12	0.771904091392439\\
13	0.777281953590168\\
14	0.778221284374065\\
15	0.7730066715847\\
16	0.772518442758351\\
17	0.771820759540536\\
18	0.774208314814909\\
19	0.774978089237367\\
20	0.762765920394484\\
};
\addlegendentry {{\Large s2}}
\addplot[color={purple}, mark={star}]
table[row sep=crcr]{%
1	0.571474875195974\\
2	0.675312231042555\\
3	0.705984121766989\\
4	0.716775804251697\\
5	0.751121363859044\\
6	0.73382704125233\\
7	0.757146966662187\\
8	0.75768102691019\\
9	0.752797640999932\\
10	0.743258047374474\\
11	0.756937001846725\\
12	0.754813862167644\\
13	0.765013420825248\\
14	0.778874823587747\\
15	0.762973448937797\\
16	0.767879598031928\\
17	0.769088959247574\\
18	0.746368272005167\\
19	0.752541820631393\\
20	0.768087777045378\\
};
\addlegendentry {{\Large s3}}
\addplot[color={brown}, mark={diamond*}]
table[row sep=crcr]{%
1	0.767604637138717\\
2	0.769929305891828\\
3	0.762157060919805\\
4	0.735767835867393\\
5	0.743512110918758\\
6	0.755833278625771\\
7	0.778664236968829\\
8	0.773203067561396\\
9	0.758610722451041\\
10	0.778752617450897\\
11	0.794174110672067\\
12	0.786078575600135\\
13	0.781627873057487\\
14	0.784584047550623\\
15	0.79282145228034\\
16	0.774800854703678\\
17	0.771372078508285\\
18	0.753077713435502\\
19	0.786568315416121\\
20	0.789581219951017\\
};
\addlegendentry {{\Large s4}}
\addplot[color={red}, mark={x}]
table[row sep=crcr]{%
1	0.335392622620581\\
2	0.332043111066769\\
3	0.334597545474621\\
4	0.337678400104026\\
5	0.332057650567736\\
6	0.329857514633496\\
7	0.329582667601162\\
8	0.330800294478031\\
9	0.325731003810904\\
10	0.321566062378603\\
11	0.329067603414451\\
12	0.333154316076968\\
13	0.332134809135495\\
14	0.341081941775255\\
15	0.336722952619112\\
16	0.334685507834207\\
17	0.336416876931158\\
18	0.327404518189118\\
19	0.330965011755791\\
20	0.333209344369545\\
};
\addlegendentry {{\large KF}}
\end{axis}
\end{tikzpicture}%